\documentclass[12pt,preprint]{aastex}
\begin{document}

\title{NON-THERMAL CONTINUUM TOWARD SGRB2(N-LMH)}

\author{J. M. Hollis\altaffilmark{1}, P. R. Jewell\altaffilmark{2}, 
Anthony J. Remijan\altaffilmark{2}, \& F. J. Lovas\altaffilmark{3}}

\altaffiltext{1}{NASA Goddard Space Flight Center, Computational and Information Sciences and Technology Office, Code 606, Greenbelt, MD  20771}
\altaffiltext{2}{National Radio Astronomy Observatory, 520 Edgemont Road, Charlottesville, 
     VA, 22903-2475}
\altaffiltext{3}{Optical Technology Division, National Institute of Standards 
and Technology, Gaithersburg, MD  20899}

\begin{abstract}

An analysis of continuum antenna temperatures observed in the Green
Bank Telescope (GBT) spectrometer bandpasses is presented for
observations toward SgrB2(N-LMH).  Since 2004, we have identified four
new prebiotic molecules toward this source by means of rotational
transitions between low energy levels; concurrently, we have observed
significant continuum in GBT spectrometer bandpasses centered at 85
different frequencies in the range of 1 to 48 GHz.  The continuum
heavily influences the molecular spectral features since we have
observed far more absorption lines than emission lines for each of
these new molecular species.  Hence, it is important to understand the
nature, distribution, and intensity of the underlying continuum in the
GBT bandpasses for the purposes of radiative transfer, i.e., the means
by which reliable molecular abundances are estimated.  We find that
the GBT spectrometer bandpass continuum is consistent with
optically-thin, non-thermal (synchrotron) emission with a flux density
spectral index of -0.7 and a Gaussian source size of $\sim$143$''$
at 1 GHz that decreases with increasing frequency as $\nu^{-0.52}$.
Some support for this model is provided by high frequency Very Large Array (VLA) observations of SgrB2.

\end{abstract}


The North source in SgrB2, i.e. SgrB2(N), is a giant star-forming
region that lies along the galactic equator near the center of our
Galaxy.  When searching for new interstellar molecules in SgrB2, the
pointing position SgrB2(N-LMH) has produced excellent results.  There
is another strong molecular source 45$''$ to the south of this
pointing position known as Main, i.e. SgrB2(M), that can influence
molecular detections toward SgrB2(N-LMH) if the telescope beam is
large.  Both SgrB2(N) and SgrB2(M) contain molecular maser emitting
spots, ultracompact HII continuum sources, compact hot molecular cores 
of arcsecond dimensions, extended HII regions, and cold extended
molecular regions of arcminute dimensions.  In addition, small-scale
and large-scale shock phenomena pervade this region (e.g., Chengalur
\& Kanekar 2003).  In particular, the hot
molecular core known as the LMH (Large Molecule Heimat) has been a
primary source to be searched for mm-wave rotational transitions
between high energy levels for species that are found in emission and 
confined to its $\sim$5$''$ diameter.   For example, from
interferometric observations the emission from high-energy transitions
of methyl formate (CH$_3$OCHO; Mehringer et al.\ 1997), acetic acid
(CH$_3$COOH; Mehringer et al.\ 1997), ethyl  cyanide (CH$_3$CH$_2$CN;
Miao \& Snyder 1997), formic acid (HCOOH; Liu, Mehringer, \& Snyder
2001), and acetone (CH$_3$COCH$_3$; Snyder et al.\ 2002) is seen
largely confined to the LMH core.  High energy transitions of methanol 
(CH$_3$OH) are also confined to the LMH hot core, and were used by
Pei, Liu, \& Snyder (2000) to derive a rotational temperature of 
170$\pm$13 K that is usually assumed to characterize other large
molecules in the core.

On the other hand, recent observations with the Green Bank Telescope
(GBT) toward SgrB2(N-LMH) indicate that the molecular halo surrounding
the LMH is a rich source of an entirely different set of large complex
molecules. In this cold halo region, transitions between low-energy
levels of large interstellar molecules tend to occur in the frequency 
range of 1 and 48 GHz.  For example, glycolaldehyde (CH$_2$OHCHO) was
first detected in this region with the GBT by means of the
1$_{10}$-1$_{01}$, 2$_{11}$-2$_{02}$, 3$_{12}$-3$_{03}$, and
4$_{13}$-4$_{04}$ rotational transitions at 13.48, 15.18, 17.98, and
22.14 GHz, respectively; all levels involved in these four transitions
have energies less than 6.5 K.  Only the 1$_{10}$-1$_{01}$ transition
of glycolaldehyde was seen solely in emission; all the other
transitions were seen predominantly in absorption.  An analysis of
these four rotational transitions yields a glycolaldehyde state
temperature of $\sim$8 K (Hollis et al.\ 2004a) which is probably 
characteristic of the cold halo region surrounding the LMH.  Other
large molecules that have been detected predominantly in absorption
toward SgrB2(N-LMH) with the GBT by means of transitions between
low-energy levels include propenal (CH$_2$CHCHO; Hollis et al.\
2004b), propanal (CH$_3$CH$_2$CHO; Hollis et al.\ 2004b), acetamide 
(CH$_3$CONH$_2$; Hollis et al.\ 2006a) and cyclopropenone
(c-H$_2$C$_3$O; Hollis et al.\ 2006b).  
It is important to note that other molecular sources observed with the
GBT such as the dark cloud TMC-1 or the IRC+10216 circumstellar nebula
show no measurable continuum in the spectrometer bandpass.  
However, low energy transitions of H$_2$CO (e.g., the 1$_{11}$-1$_{10}$)
in TMC-1 have been observed to exhibit absorption against the cosmic
background (Palmer et al. 1969).  It is also noteworthy that
the GBT detection of ketenimine (CH$_2$CNH; Lovas et al.\ 2006) in
absorption toward SgrB2(N-LMH) suggests that an intermediate temperature region is
responsible because the transitions involved are between intermediate
energy levels that range from 33 to 51 K.

Since we have observed far more absorption lines than emission lines
of new interstellar molecules in the cold halo toward SgrB2(N-LMH), it
is important to understand the nature of the global continuum in the
GBT spectrometer bandpass.  As a consequence of our deep integration
observations in search of new molecules, extensive low noise level
data have been collected in 85 spectrometer bandpasses centered at
different frequencies in the range of 1 to 48 GHz.  The continuum
antenna temperature in each spectrometer bandpass is observed to be a
function of bandpass center frequency, thus providing the means for 
analyzing the source of continuum reported here.

Spectral line observations over the range of 1 to 48 GHz were conducted with 
the NRAO\footnote{The National Radio Astronomy Observatory is a facility of the 
National Science Foundation, operated under cooperative agreement by Associated
Universities, Inc.} 100-m Robert C. Byrd Green Bank Telescope (GBT) from 2004 
March 4 to 2005 November 12.  The GBT spectrometer was configured to 
provide four intermediate frequency (IF) bandwidths at a time in two 
polarizations through the use of offset oscillators in the IF.  Table 1 lists 
the receiver band, the receiver tuning range, the spectrometer bandwidth per 
IF, the spectrometer channel spacing, and 2004 and 2005 observation dates in 
the first six columns.  GBT half-power beamwidths can be approximated by:

\begin{equation}
\theta_B \approx \frac{740''}{\nu}
\end{equation}

\noindent
where the observing frequency ($\nu$) is in units of GHz.  Over the course of 
the observations, the spectrometer bandpasses were centered at a total of 85 
different frequencies in the range of 1.35 GHz to 47.68 GHz which correspond 
to $\theta_B\approx$ 548$"$ and $\theta_B\approx$ 16$"$, respectively.  The 
SgrB2(N-LMH) J2000 pointing position employed was $\alpha$ =
17$^h$47$^m$19$^s$.8, $\delta$ = -28$^o$22$'$17$"$ which correspond to galactic coordinates of $\ell$ = 0.$^o$6775, b = -0.$^o$0271. An LSR source
velocity of +64 km s$^{-1}$ was assumed.  Spectrometer data 
were taken in the OFF-ON position-switching mode, with the OFF position 60$'$ 
East in azimuth with respect to the ON source position.   A single scan 
consisted of 2 minutes in the OFF source position followed by 2 minutes in the 
ON source position.  SgrB2(N-LMH) was observed in this manner above 10$^o$ 
elevation from source rise to source set (i.e., a six-hour track).  Note that at ON source transit, the OFF source position in galactic coordinates is $\ell$ = 1.$^o$205 and b = -0.$^o$876, and all OFF positions along the six-hour track are well clear of sources of contamination.  The accumulated scans over two or three tracks for the two polarization outputs from the 
spectrometer were averaged in the final data reduction process.  The antenna temperatures of the spectral line and continuum 
emission produced in the spectrometer bandpass are on the T$_A^*$ scale 
(Ulich \& Haas 1976) with estimated 20\% uncertainties.

The instrumental slope in a GBT spectrometer bandpass can be quite significant 
in the presence of source continuum emission (see Figure 3 of Hollis 2006), and 
there can be unpredictable effects at the edges of the bandpass.  Therefore, a 
continuum antenna temperature (T$_C$) was estimated at or near the center of
each of the 85 different spectrometer bandpasses.  Figure 1(a) is a
linear-linear plot of T$_C$ versus spectrometer bandpass center
frequency and Figure 1(b) is a log-log plot of the same data.  The
-1.06 slope of Figure 1(b) was determined by a linear least squares
fit, and represents the spectral index of T$_C$.  The fit result is
superimposed as a dotted line on Figure 1(a).  These plots show that
T$_C$ is a highly predictable function of frequency for GBT
spectrometer bandpasses in the range of 1 to 48 GHz.  The main beam
brightness temperature (T$_B$) is determined from T$_C$ divided by the
GBT beam efficiency which can be estimated by:

\begin{equation}
\eta_B=-15.52\times10^{-5}\nu^2 -22.59\times10^{-4}\nu + 0.98
\end{equation}

\noindent
where $\nu$ is in the range of 1.35 GHz to 47.68 GHz which corresponds to
$\eta_B \approx$ 0.97 and $\eta_B \approx$ 0.53, respectively.
Equation (2) derives from a fit to the Ruze (1966) formulation,
assuming a GBT surface accuracy of 390 microns and an aperture
efficiency $\eta_A$($\nu$=0) = 0.71, and favorably compares with
similar results in Langston \& Turner (2007).  Figure 2(a) is a
log-log plot of T$_B$ as a function of the center frequency in the 85
spectrometer bandpasses. The -0.94 slope of Figure 2(a) was determined
by a linear least squares fit, and represents the spectral index of
T$_B$.  In what follows, we develop a method for estimating the
spectral index of the flux density ($S$) of the observed continuum that
influences the bandpass shape.  The value of the spectral index of $S$
is an indication of the nature of the continuum emission itself.

A spectral index ($\alpha$) of a source can be cast in terms of its continuum 
antenna temperature (T$_C$), or its brightness temperature (T$_B$), or its 
flux density ($S$) each of which has a value proportional to $\nu^{\alpha}$.  
The spectral indices for T$_C$, T$_B$, and $S$ will, in general, have
different numerical values, and are related through the following
equation for the flux density of an unmapped source:  

\begin{equation}
S = 2 k \frac{\nu^2}{c^2} \frac{\pi \theta_B^2}{4 ln(2)} T_B [1+\theta_S^2 / \theta_B^2]
\end{equation}

\noindent
where k is the Boltzmann constant, c is the speed of light, $\theta_B$
is the Gaussian half-power beam width, $\theta_S$ is an intrinsic
Gaussian source size, $\pi \theta_B^2/4ln(2)$ is the main beam solid
angle for a Gaussian pattern, and the bracketed term accounts for the
source extending beyond the telescope beam ($\theta_S > \theta_B$).
Note that the bracketed term approaches unity when the source is much
smaller than the beam and the telescope measures the total flux
density of the source.  To calculate the flux density as a
function of frequency in equation (3), we must have an independent
measure of the source size which may also be a function of frequency.

To estimate the intrinsic, deconvolved source size
($\theta_S$), we used Very Large Array (VLA) archival data taken at
observing frequencies of 4.86, 22.15, and 43.75 GHz toward the
molecular pointing position SgrB2(N-LMH).  It was necessary to
taper these VLA data in the MIRIAD task INVERT to approximate
GBT beam sizes of 152$"$, 33$''$, and 17$"$ at 4.86, 22.15, and 43.75
GHz, respectively.  The "true" source size ($\theta_S$) is assumed to be related to the observer source size ($\theta_{OBS}$) and the VLA beam ($\theta_{VLA}$) by: 

\begin{equation}
\theta_{OBS}(maj)\times\theta_{OBS}(min) = \theta_{VLA}(maj)\times\theta_{VLA}(min) + \theta_S^2
\end{equation}

\noindent
We note that despite tapering the VLA visibility data, some diffuse emission may still be resolved out, introducing an unknown systematic uncertainty into equation (4). The VLA results in Table 2 lists the observing frequency, the taper
(i.e., the fwhm parameter set in the MIRIAD task INVERT) used to
approximate the GBT beam, the convolved source size from a Gaussian
fit, the VLA restoring beam size, the source size resulting from
equation (4), and the SgrB2 component detected.  The most obvious
result from Table 2 is that the larger telescope beam at 4.86 GHz
fails to resolve the two major components of continuum at the SgrB2(N) 
and SgrB2(M) positions.  The left side of Figure 3 shows the observed
source size overlaid on a gray-scale image of the 4.86 GHz continuum;
the right side shows the deconvolved source size within the 152$"$
diameter GBT beam.  By comparison, Figure 4 shows the results for
22.15 GHz in which the two major components SgrB2(N) and SgrB2(M) are resolved.

The intrinsic source size for the North component decreases with
increasing frequency as shown in Table 2.  Based on the 22.15 and
43.75 GHz entries in Table 2, we calculate that  $\theta_S$ (North
only) = $77'' \nu^{-0.52}$ where the observed frequency ($\nu$) is in
units of GHz.   If the same frequency dependence is assumed for all 
of the continuum encompassed in larger telescope beams (e.g., at 4.86
GHz), then the following frequency relation for the effective size of 
all sources of continuum is obtained:

\begin{equation}
\theta_{S} \approx 143'' (\nu)^{-0.52}
\end{equation}

\noindent
which yields $\theta_S$ = 62.8$''$ for an observing frequency of 4.86
GHz in agreement with the corresponding Table 2 entry in which the
continuum components are unresolved.  Comparison of the North only
source size and equation (5) for the effective size of unresolved
components are consistent, for example, at 4.86 GHz when one considers 
that SgrB2(N) and SgrB2(M) probably have similar continuum and are
spatially separated by $\approx$45$''$.

Using the brightness temperature data from Figure 2(a) and the
estimate of the source size given by equation (5), the flux density
$S$ as a function of frequency is computed by means of equation (3).
Figure 2(b) is a log-log plot of $S$.  The -0.7 slope of Figure
2(b) was determined by a linear least squares fit, and represents the
spectral index of $S$.  Since the slope of Figure 2(b) is negative and
quite linear, showing no indication of turnover at lower frequency,
these data are consistent with a synchrotron spectrum from a power-law 
distribution of electrons in the optically thin case (see Figure 6.12
of Rybicki \& Lightman 1979).

When the GBT is pointed toward SgrB2(N-LMH), the 1 to 48 GHz continuum 
detected in the spectrometer bandpass is consistent with an
optically-thin, non-thermal source.  Moreover, this pointing position
is toward the K2 ultracompact HII region (see $\S$2) with other
thermal sources in its vicinity; these smaller thermal sources would
be severely beam diluted in the present GBT observations.  It has been
long known that there is a continuous distribution of
background continuum radiation from our Galaxy composed
of a thermal spectrum and a
non-thermal spectrum with maximum intensity toward the plane (e.g., Kraus 1966).  Moreover, Yusef-Zadeh et al.\ (2003) have reported non-thermal emission
specifically near the galactic center, and Sofue (1994) has argued
that the radio continuum from the 3$^o$ x 3$^o$ galactic center region
is a mixture of synchrotron and free-free emissions.  Thus, the
results in this work are consistent with a mixture of thermal and
non-thermal radio emission expected since SgrB2 lies along the
galactic equator near the galactic center.   Further, non-thermal emission toward SgrB2 at frequencies lower than 1.5 GHz has been observed (F. Yusef-Zadeh 2007, private communication).  Moreover, Yusef-Zadeh et al.\ (2007) report a low-energy cosmic ray model that predicts a non-thermal radio spectrum and Crocker et al.\ (2007) suggest that secondary electrons from gamma rays could produce non-thermal radio emission in SgrB2.   

In particular, SgrB2(N) and its many sources of continuum emission
have been previously well-studied with the Very Large Array (VLA),
albeit at higher spatial resolution than the GBT spectrometer data
presented here.  For example, Gaume et al.\ (1995, Figure 10) produced
a 2.7$''$ resolution spectral index image of the flux density from
continuum images at 1.5 GHz and 22.3 GHz.  The
locations of ultracompact HII regions show up prominently with a
spectral index of the flux density  $\sim$1 (i.e., the slope), indicating
optically-thick thermal emission from dust and/or emission primarily
from free-free radiation.
Similarly, Mehringer \& Menten (1997, Figure 2) produced a $\sim$3$''$
resolution spectral index image from 8.4 GHz and 44 GHz data, showing
that the spectral index of the flux density is $\sim$0.8 toward the
ultracompact HII region K2; however, northeast of K2 (at $\sim$3$''$
resolution and over an area with a scale size of $\sim$20$''$) the
spectral index of the flux density is largely zero, indicating
optically-thin thermal emission of an extended but clumpy HII region.  
Moreover, Mehringer \& Menten (1997) obtain a 44 GHz thermal continuum
flux density of 5.5 Jy for SgrB2(N) from high-resolution VLA
observations while we obtain 4.6 Jy at 44 GHz for the non-thermal
continuum flux density in low-resolution GBT observations reported
here.  These flux density results are disparate and mutually exclusive
because of missing short antenna spacings in the VLA observations that 
are needed to detect the large-scale spatial component that the GBT samples well.  It is interesting to note that Akabane et al.\ (1988) compared matched $\sim$40$''$ resolution observations of the Effelsberg 100-m telescope at 23 GHz and the Nobeyama 45-m telescope at 43 GHz to conclude that SgrB2 is largely thermal.  At both frequencies, they obtain an observed source scale size of $\sim$45$''$ toward both SgrB2(N) and SgrB2(M), suggesting that the source size is not a function of frequency.  Thus, it is likely that the Akabane et al. (1988) methodology samples different gas than that gas sampled by the GBT spectrometer at 85 different frequencies, further confirming that disentangling thermal and non-thermal emission is often a vexing problem. 

In summary, an analysis of continuum antenna temperatures detected in
the GBT spectrometer bandpasses is presented
for observations toward SgrB2(N-LMH).  The continuum controls the
absorption features seen in molecular absorption transitions between
low energy levels of several new complex molecules observed with the
GBT, and therefore influences estimates of molecular abundances.  The
analysis herein assesses the nature and the effective scale size of the continuum
source observed within the GBT spectrometer bandpasses centered on
different frequencies.  The methodology employed determines the spectral indices of
the continuum antenna temperature and the brightness temperature
across all spectrometer bandpasses, determines the source size as a
function of frequency from archival VLA data, and then estimates the
spectral index of the flux density.  As a result, the GBT spectrometer
bandpass continuum seen toward SgrB2(N-LMH) is consistent with
optically-thin, non-thermal (synchrotron) emission with a flux density
spectral index of -0.7 and a Gaussian source size of $\sim$143$''$ at
1 GHz that decreases with increasing frequency as $\nu^{-0.52}$.  We thank Ed Churchwell, Farhad Yusef-Zadeh, Michael Remijan, and an anonymous reviewer for helpful comments.

\clearpage

\begin{deluxetable}{lccccc}
\tabletypesize{\scriptsize}
\tablewidth{40pc}
\tablecolumns{6}
\tablecaption{Observational Parameters}
\tablehead{
\colhead{Band} & \colhead{Receiver} & \colhead{IF Bandwidth} & \colhead{Resolution} & \colhead{2004} & \colhead{2005}\\
\colhead{} & \colhead{(GHz)} & \colhead{(MHz)} & \colhead{(kHz)} & \colhead{} & \colhead{}\\
\colhead{(1)} & \colhead{(2)} & \colhead{(3)} & \colhead{(4)} & \colhead{(5)} & \colhead{(6)}\\
}
\startdata
L & 1.15-1.73 & 50 & 6.1 & & Sept 9\\
S & 1.73-2.60 & 50 & 6.1 & & Sept 8\\
C & 3.95-5.85 & 200 & 24.4 & & Sept 28; Oct 9-11\\
X & 8.00-10.1 & 200 & 24.4 & & Sept 6, 18, 19; Nov 10\\
Ku & 12.00-15.4 & 200 & 24.4 & Mar 4; Apr 17 & Sept 7, 14, 22\\
K & 18.00-22.4 & 200 & 24.4 & Feb 25, 29; Mar 13, 29; Apr 5, 6, 15 & Apr 1; Nov 12\\
K & 22.00-26.5 & 200 & 24.4 & Mar 10; Apr 16 & Oct 11-14\\
Q & 40.00-48.0 & 800 & 390.7 & & Mar 14, 15, 19, 22, 30, 31; Apr 5\\
\enddata
\end{deluxetable}

\clearpage

\begin{deluxetable}{lccccc}
\tabletypesize{\scriptsize}
\tablewidth{27pc}
\tablecolumns{6}
\tablecaption{Continuum Source Size ($\theta_S$) Estimates\tablenotemark{a}}
\tablehead{
\colhead{} & \colhead{Taper} & \colhead{Convolved Source} &
\colhead{VLA Beam} & \colhead{} & \colhead{}\\
\colhead{Frequency} & \colhead{FWHM} &
\colhead{$\theta(maj)\times\theta(min)$} &
\colhead{$\theta(maj)\times\theta(min)$} & \colhead{$\theta_S$} & \colhead{SgrB2}\\
\colhead{(GHz)} & \colhead{$''$} & \colhead{$''\times''$} &
\colhead{$''\times''$} & \colhead{$''$} & \colhead{Component}\\
\colhead{(1)} & \colhead{(2)} & \colhead{(3)} & \colhead{(4)} &
\colhead{(5)} & \colhead{(6)}\\
}
\startdata
4.86 & 152 & 215.3$\times$101.9 & 150.2$\times$119.5 & 63.2 & North+Main\\
     &  &  &  &  & \\
22.15 & 33 & 36.8$\times$30.5 & 33.4$\times$26.4 & 15.5 & North\\
      &    & 36.9$\times$30.1 &                  & 15.1 & Main\\
     &  &  &  &  & \\
43.75 & 17 & 26.4$\times$13.7 & 23.1$\times$10.5 & 10.9 & North\\
      &    & 19.1$\times$7.6  &                  & ... & Main\\
\enddata
\tablenotetext{a}{Derived from VLA restoring beams that approximate corresponding GBT beam sizes (see text and equation 4).}

\end{deluxetable}

\clearpage

\begin{figure}
\plotone{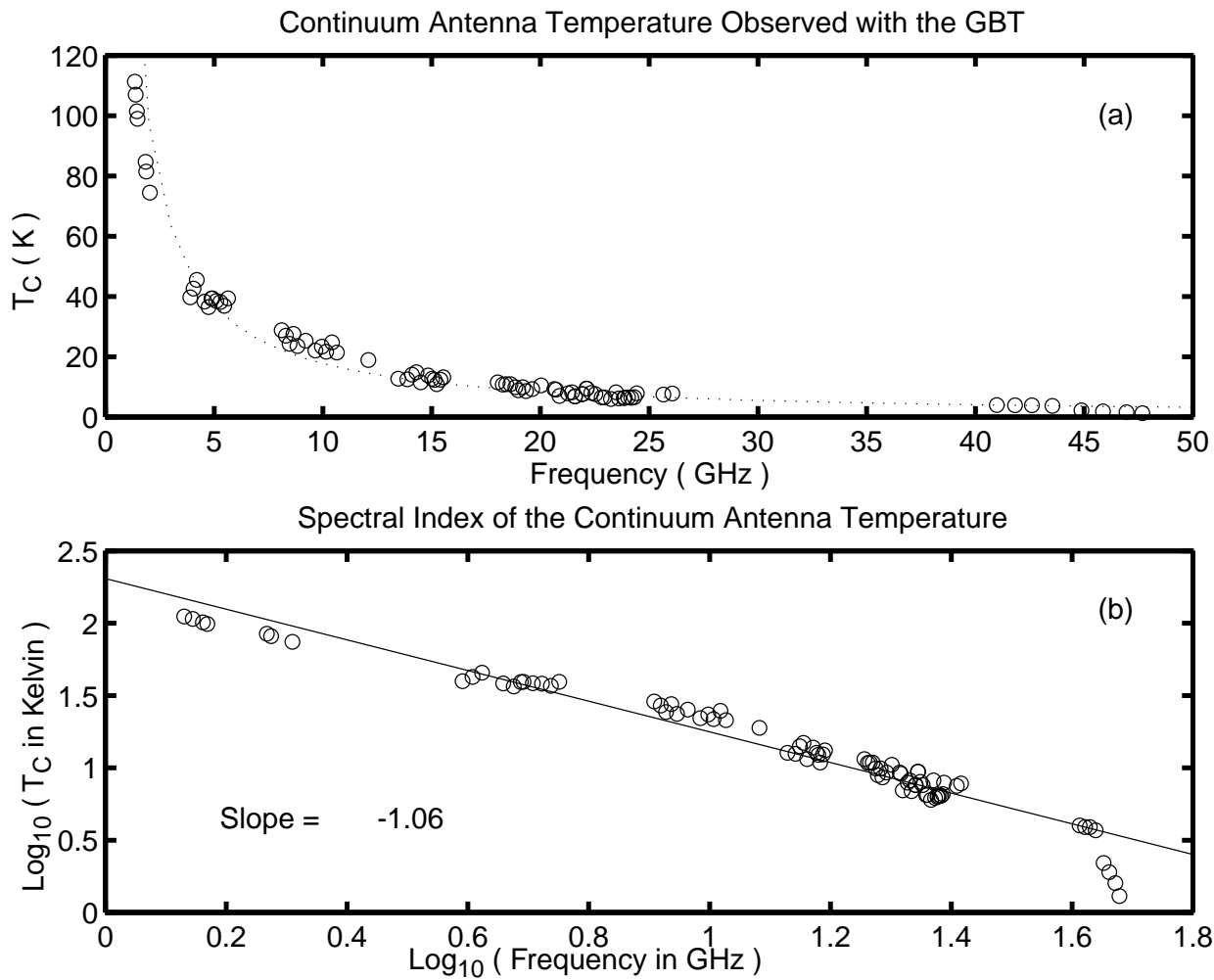}
\caption{Continuum antenna temperatures observed with the GBT
  spectrometer toward SgrB2(N-LMH).  Panel (a) plots continuum antenna
  temperature as a function of spectrometer bandpass center frequency.
  Panel (b) is a log-log plot of panel (a) data.  A linear least
  squares fit to panel (b) data yields a slope of -1.06 which
  determines the spectral index of the continuum antenna temperature.
  The dotted line in panel (a) results from the fit to data in panel (b).}
\end{figure}

\clearpage

\begin{figure}
\plotone{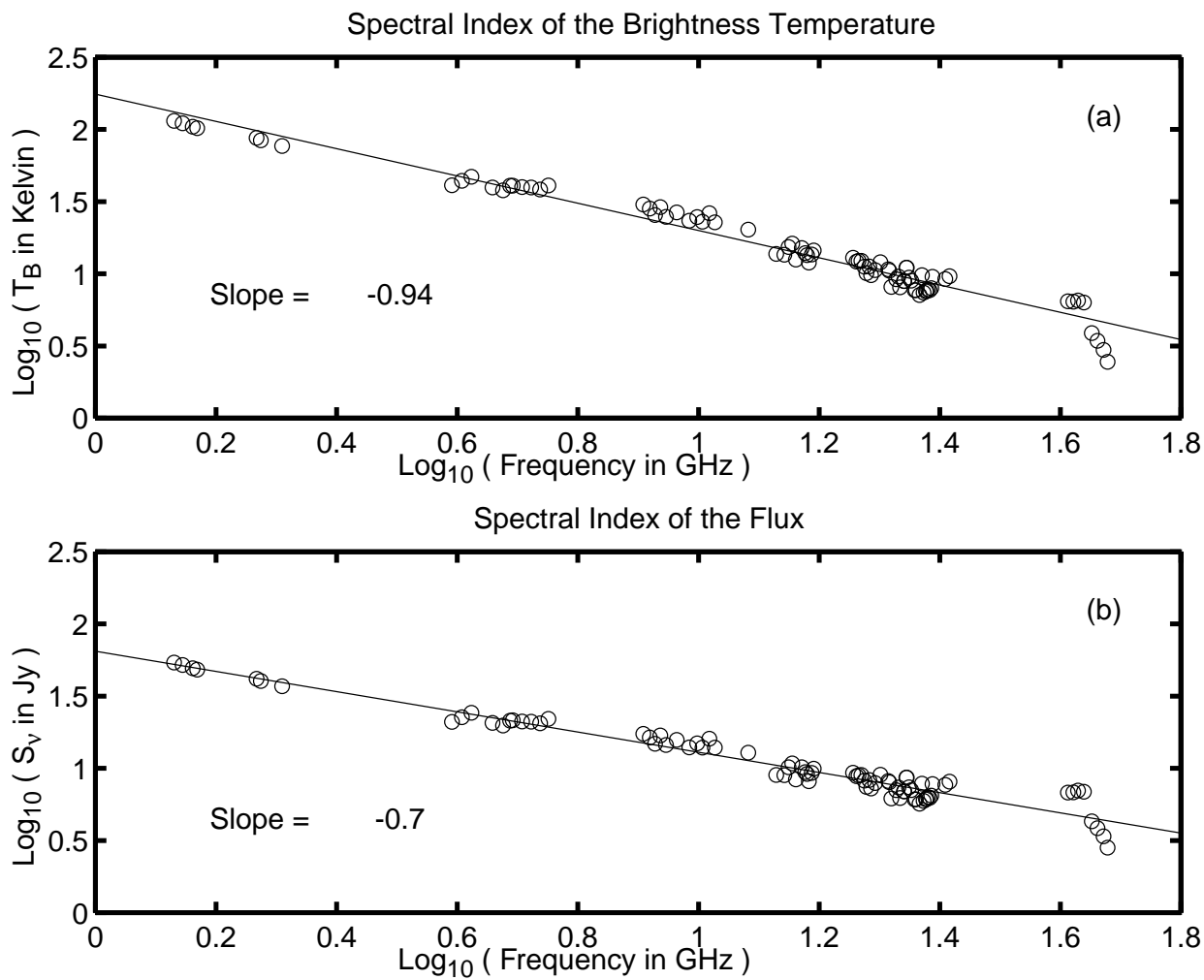}
\caption{Spectral indices of the brightness temperature and flux
  density observed with the GBT spectrometer toward SgrB2(N-LMH).
  Panel (a) is a log-log plot of the brightness temperature as a
  function of bandpass center frequency.  The resultant slope of -0.94
  determines the spectral index of the brightness temperature.   Panel
  (b) is a log-log plot of the observed flux density as a function of
  bandpass center frequency which is dependant upon the observed source
  size (see text).  The resultant slope of -0.7 determines the
  spectral index of the flux density, suggesting that the source of
  continuum is optically thin and non-thermal.}
\end{figure}

\clearpage

\begin{figure}
\epsscale{0.55}
\plotone{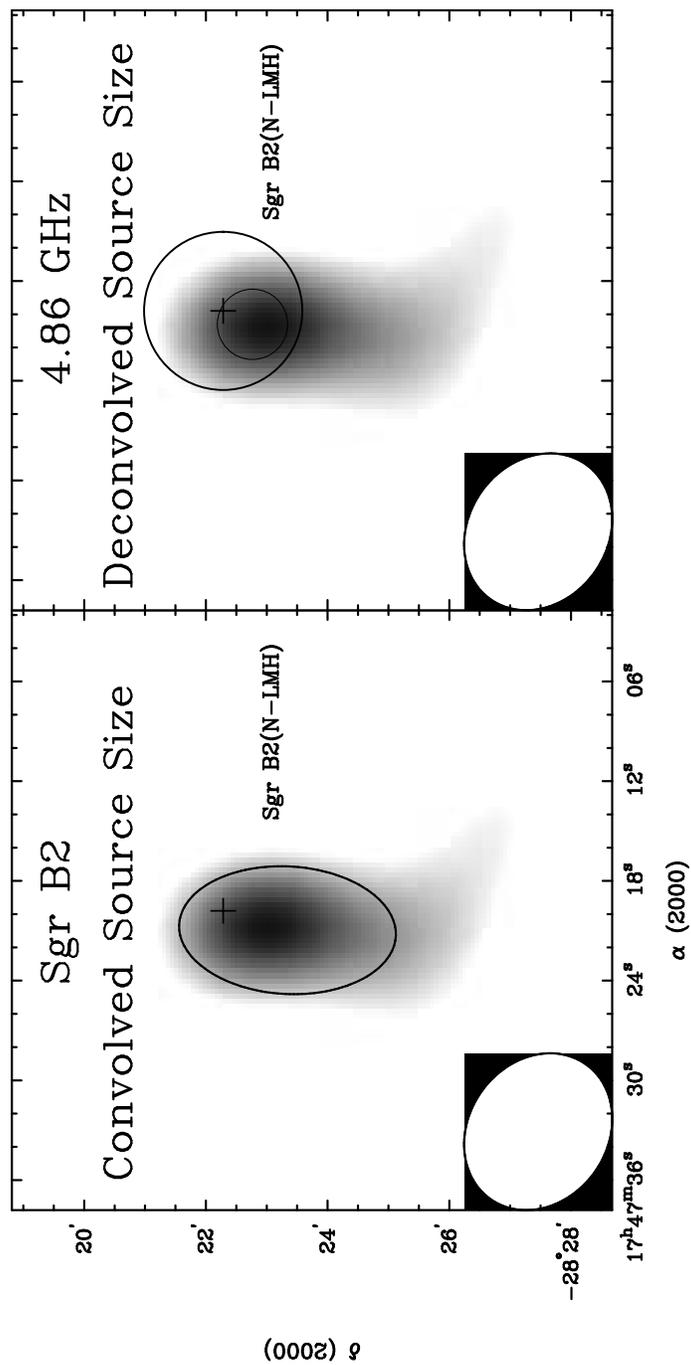}
\caption{VLA observations of SgrB2 at 4.86 GHz.  These VLA
  observations were processed to approximate the GBT beam (see text
  and Table 2).  The left side shows the observed source size overlaid
  on a gray-scale image of the 4.86 GHz continuum; the right side
  shows the deconvolved source size within the 152$''$ diameter GBT
  beam.  Note that the primary sources of continuum (i.e., SgrB2(N)
  and SgrB2(M)) are unresolved.}
\end{figure}

\clearpage

\begin{figure}
\epsscale{0.8}
\plotone{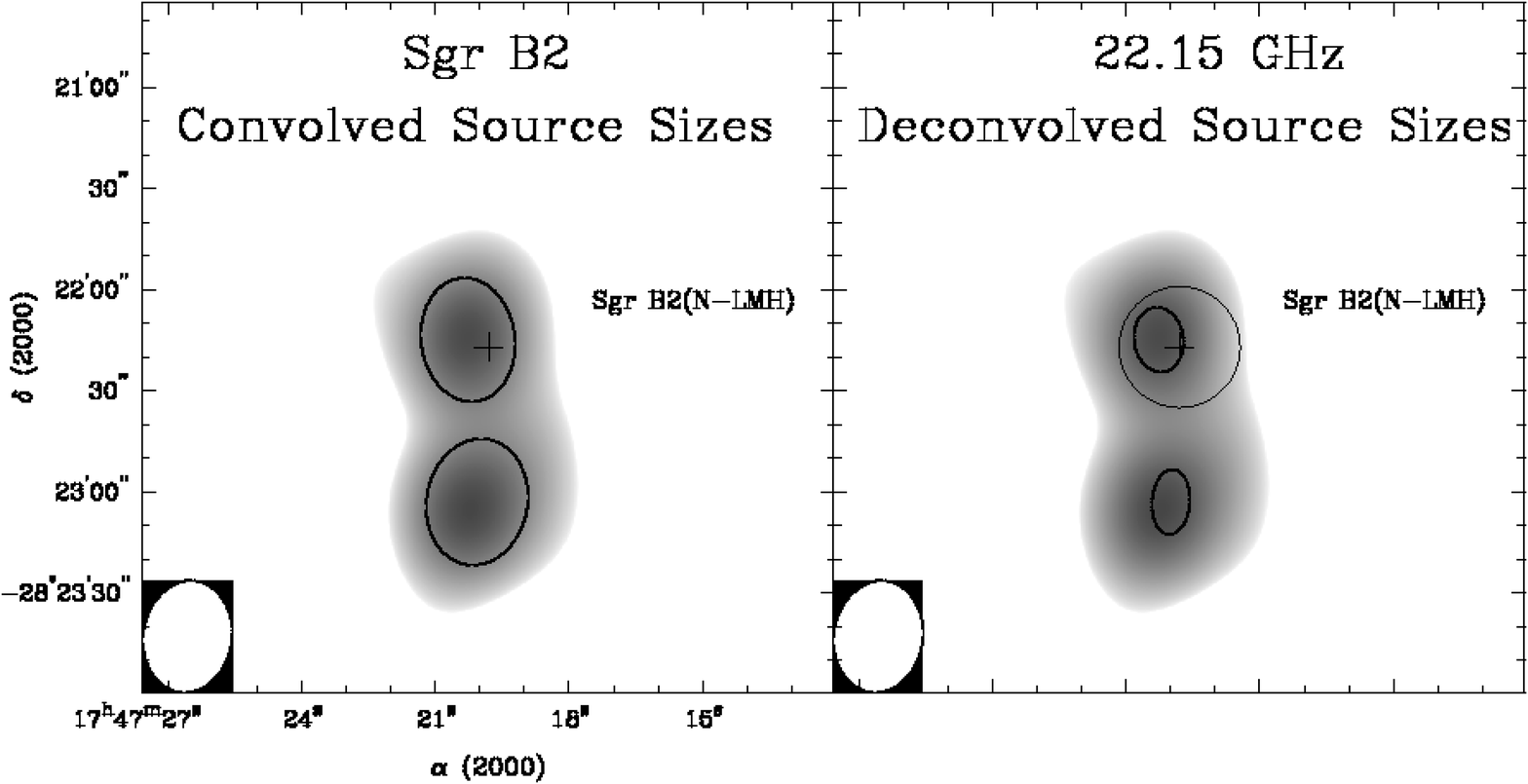}
\caption{VLA observations of SgrB2 at 22.15 GHz. These VLA
  observations were processed to approximate the GBT beam (see text
  and Table 2).  The left side shows the observed source size overlaid
  on a gray-scale image of the 22.15 GHz continuum; the right side
  shows the deconvolved source size within the 33$''$ diameter GBT
  beam.  Note that the primary sources of continuum (i.e., SgrB2(N) and SgrB2(M)) are resolved.}
\end{figure}

\end{document}